\documentclass[aps,prl,twocolumn,superscriptaddress]{revtex4-1}
\usepackage{blindtext} \usepackage{amsmath,amssymb}
\usepackage{graphicx} \usepackage{amsmath} \usepackage{float}
\usepackage{color} \usepackage{upgreek}
\usepackage{lipsum}

\begin{document} \title{Experimental Realisation of a Thermal Squeezed
State of Levitated Optomechanics}

\author{Muddassar Rashid} \affiliation{Department of Physics \&
Astronomy, University of
  Southampton, SO17 1BJ, UK}

\author{ Tommaso Tufarelli} \email{Tommaso.Tufarelli@nottingham.ac.uk}
\affiliation{School of Mathematical Sciences, University of
Nottingham,
  NG7 2RD, UK}

\author{James Bateman} \email{j.e.bateman@swansea.ac.uk}
\affiliation{Department of Physics \& Astronomy, University of
Southampton, SO17 1BJ, UK} \affiliation{Department of Physics, Swansea
University, SA2 8PP, UK}
 
\author{Jamie Vovrosh} \affiliation{Department of Physics \&
Astronomy, University of
  Southampton, SO17 1BJ, UK}

\author{David Hempston} \affiliation{Department of Physics \&
Astronomy, University of
  Southampton, SO17 1BJ, UK}

\author{M. S. Kim} \email{m.kim@imperial.ac.uk} \affiliation{QOLS,
Blackett Laboratory, Imperial College London, London,
  SW7 2AZ, UK}

\author{Hendrik Ulbricht} \email{h.ulbricht@soton.ac.uk}
\affiliation{Department of Physics \& Astronomy, University of
  Southampton, SO17 1BJ, UK}

\begin{abstract}
We experimentally squeeze the thermal motional state of an optically
levitated nanosphere, by fast switching between two trapping
frequencies. The measured phase space distribution of the
center-of-mass of our particle shows the typical shape of a squeezed
thermal state, from which we infer up to 2.7 dB of squeezing along one
motional direction. In these experiments the average thermal occupancy
is high and even after squeezing the motional state remains in the
remit of classical statistical mechanics. Nevertheless, we argue that
the manipulation scheme described here could be used to achieve
squeezing in the quantum regime, if preceded by cooling
of the levitated mechanical oscillator. Additionally, a higher degree
of squeezing could in principle be achieved by repeating the
frequency-switching protocol multiple times. \end{abstract}

\maketitle

While squeezing a quantum state of light \cite{Lvovsky2014} has a long
history of experiments, the squeezing of a massive mechanical harmonic
oscillator has so far not seen many experimental realisations. The
first demonstration of squeezing in a classical mechanical oscillator
was by Rugar et.al~\cite{Rugar1991}. Squeezing of classical motional
states in electromechanical devices by parametric amplification and
weak measurements has been subsequently
proposed~\cite{szorkovszky2011mechanical}, and experimentally
demonstrated in an optomechanical system~\cite{pontin2014squeezing}.
Schemes relying on sinusoidal modulation of the spring constant have
also been proposed and discussed by numerous authors
\cite{Mari2009,Farace2012,Woolley2008,serafini2009generation}. In
optomechanical cavities \citeauthor{Genoni2015} suggested that
squeezing below the ground-state fluctuations (quantum squeezing for brevity) may be attainable via continuous measurements and feedback \cite{Genoni2015}. Quantum squeezing of a high-frequency mechanical oscillator has only been
experimentally demonstrated very recently, in a microwave
optomechanical device~\cite{wollman2015quantum,Pirkkalainen2015}. Also
only very recently a hybrid photonic-phononic waveguide device has
shown the correlation properties of optomechanical two-mode
squeezing~\cite{riedinger2016non}.
Another interesting method of generating squeezing, of relevance to
this Letter, relies on non-adiabatic shifts of the mechanical
frequency. Such method was initially discussed in relation to light
fields \cite{janszky1986squeezing,lo1990squeezing}. Similar ideas, utilising impulse
kicks on a mechanical oscillator, have been discussed
\cite{Asjad2014,alonso2016generation}.
In this Letter we report the first experimental demonstration of mechanical squeezing via non-adiabatic frequency shifts, thus realising a useful
tool to manipulate the state of a levitated optomechanical system.
\begin{figure*}[t!]
  \centering
 \includegraphics[width=1\textwidth]{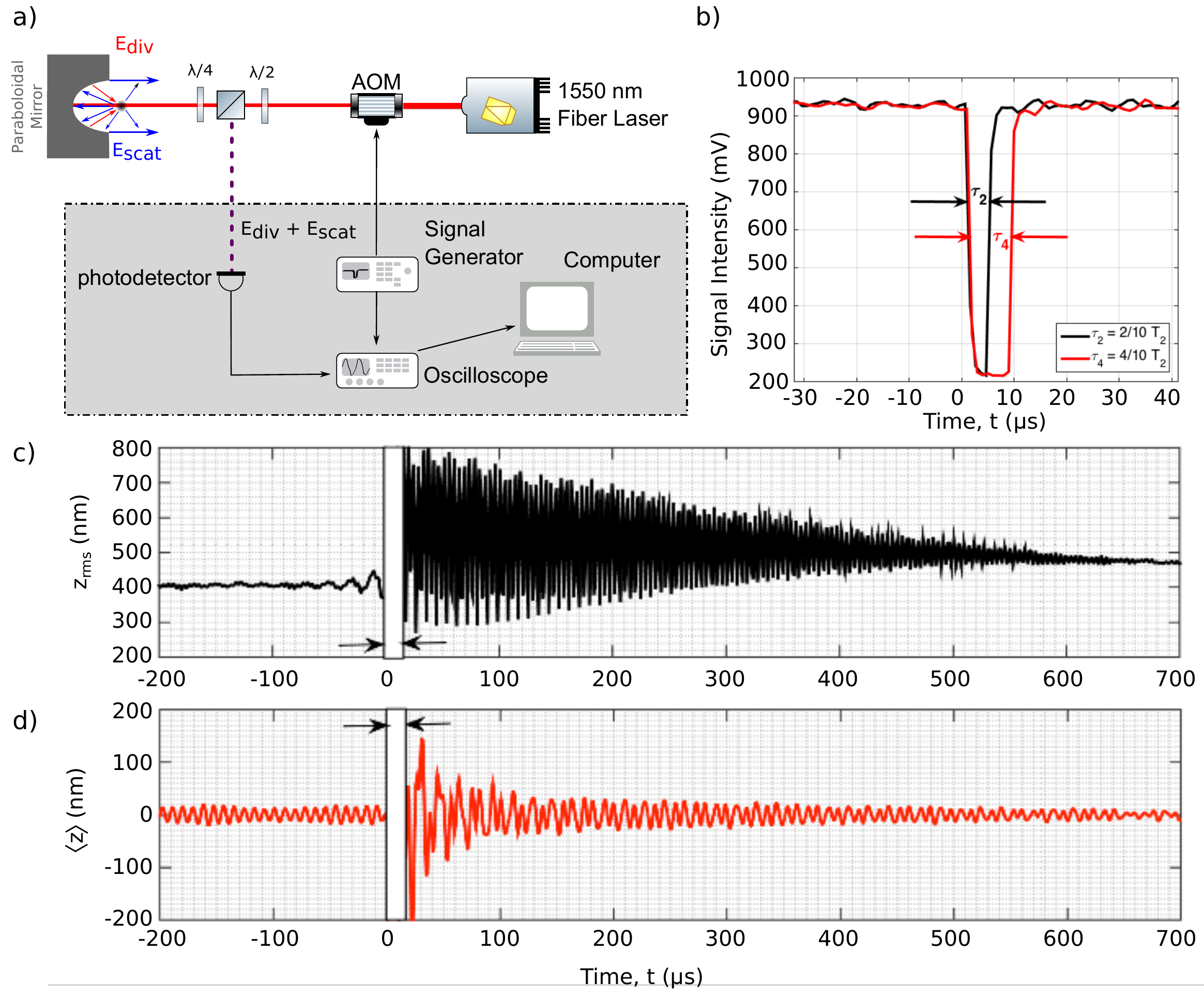}
 \caption{{\bf Experimental implementation of squeezing levitated
     optomechanics.} {\bf a}) Schematic of the squeezing setup. The
   paraboloidal trapping system demonstrates the $E_{div}$, divergence
   field and the $E_{scat}$, Rayleigh scattered field from the trapped
   particle. The grey region is for homodyne detection, as well as,
   pulse application. {\bf b}) Negative square pulse for squeezing
   generation, as seen by the photodetector. Two different pulse
   durations are shown, $\tau_2$ and $\tau_4$. {\bf c}) Root mean
   square position as a function of time,
   $z_{\rm rms}(t)=\sqrt{\langle (z-\langle z\rangle)^2\rangle}$,
   obtained from 1500 pulse sequences applied to the same
   particle. Oscillations for $t<0$ are due to band pass
   filtering. {\bf d}) Time-dependence of the mean position
   $\langle z\rangle$ (center of the thermal distribution).  This
   quantity also shows oscillations at $\omega_1$ after the squeezing
   pulse. }
  \label{fig:one} \end{figure*}

{\it Theory--} In what follows we shall present a quantum mechanical
treatment of our squeezing protocol, in anticipation of future
experiments that may achieve quantum squeezing. Due
to linearity of the Heisenberg equations of our system, it should be
pointed out that formally identical results may be obtained through
classical statistical mechanics~\cite{supplement}. We consider a
nanosphere of mass $m$ trapped in a harmonic potential. Along the $z$
axis, we can manipulate the system by switching between two
Hamiltonians $\hat H_1,\hat H_2$, where $\hat H_j=\frac{\hat
p^2}{2m}+\frac12m\omega_j^2\hat z^2$, $\hat z,\hat p$ denote the
$z$-components of the canonical position and momentum operators, and
the trapping frequency may assume two distinct values: $\omega_1$ or
$\omega_2$ (In our experiment, we adopt $\omega_2<\omega_1$). As we shall see shortly, our squeezing protocol
relies on the rapid (i.e. \textit{non-adiabatic}) switching between
the two Hamiltonians \cite{lo1990squeezing, janszky1986squeezing}. It
is instructive to write down the annihilation operators, say $\hat a$
and $\hat b$, corresponding to the two trap frequencies ($\hbar=1$):
\begin{equation}
  \label{eq:1}
  \hat a = \sqrt{\frac{m \omega_1}{2}}\Big(\hat{z} +
    \frac{i\hat{p}}{m\omega_1} \Big), \quad
 \hat b = \sqrt{\frac{m \omega_2}{2}}\Big(\hat{z} +
    \frac{i\hat{p}}{m\omega_2}\Big). 
\end{equation} 
Through simple
algebra one may notice that $\hat a$ and $\hat b$ are related by a
squeezing transformation of the form $\hat b = \cosh(r) \hat a -
\sinh(r)\hat a^{\dagger}$, with $r \equiv \frac{1}{2}
\log(\frac{\omega_2}{\omega_1})$ the squeezing parameter. We may
exploit the mathematical relationship between modes $\hat a$ and $\hat
b$ to generate mechanical squeezing, as follows. Let the particle be
initially prepared in an arbitrary state (in our experiment, this will
be a thermal state of $\hat H_1$). At time $t=0$ we suddenly change
the trapping frequency from $\omega_1$ to $\omega_2$, such that the
Hamiltonian becomes $\hat H_2$. We then let the system evolve until a
time $t=\tau$ (the {\it squeezing pulse duration}), before rapidly
switching back to Hamiltonian $\hat H_1$. In the Heisenberg picture,
this amounts to a simple harmonic evolution $\hat b\to \hat b
e^{-i\omega_2 \tau}$ for the operator $\hat b$. In terms of the
quadratures $\hat X=(\hat a+\hat a^\dagger)/\sqrt{2}, \hat P=-i(\hat
a-\hat a^\dagger)/{\sqrt{2}}$, however, the transformation is
nontrivial: $(\hat X,\hat P)^{\intercal}\to M(\hat X,\hat
P)^{\intercal}$, where the matrix \begin{align} M&=\begin{pmatrix}
\cos(\omega_2\tau)&e^{2r}\sin(\omega_2\tau)\\
-e^{-2r}\sin(\omega_2\tau)&\cos(\omega_2\tau) \end{pmatrix}
\end{align} embodies a combination of rotation and squeezing in the
phase space of mode $\hat a$. Note that, in general, the squeezed
quadrature will be a linear combination of $\hat X$ and $\hat P$. The
associated squeezing parameter $\lambda(\tau)$ is encoded in the
singular values of $M$, and can be found as follows. Since $\det(MM^\intercal)=1$, we can parametrize the eigenvalues of
$MM^\intercal$ as $(\mu,1/\mu)$ for some parameter $\mu>0$. Note that
$\sqrt\mu$ quantifies the deformation of the \textit{standard deviations} of the
rotated quadratures. The mechanical squeezing parameter
thus reads (in dB units)
\begin{equation} \label{eq:lam}
\lambda(\tau)=10\left|\log_{10}(\sqrt\mu)\right|.
\end{equation} 
The
analytical expression for $\lambda(\tau)$ is unwieldy if $\tau$ is
left generic. It is however readily verified that maximum squeezing
can be obtained by setting $\omega_2\tau=\frac{\pi}{2}$, in which case
$\lambda_{\sf max}=10\log_{10}(\omega_1/\omega_2)$.\\ {\it
Experiments--} \begin{figure*}[t!]
  \centering
  \includegraphics[width=1\textwidth]{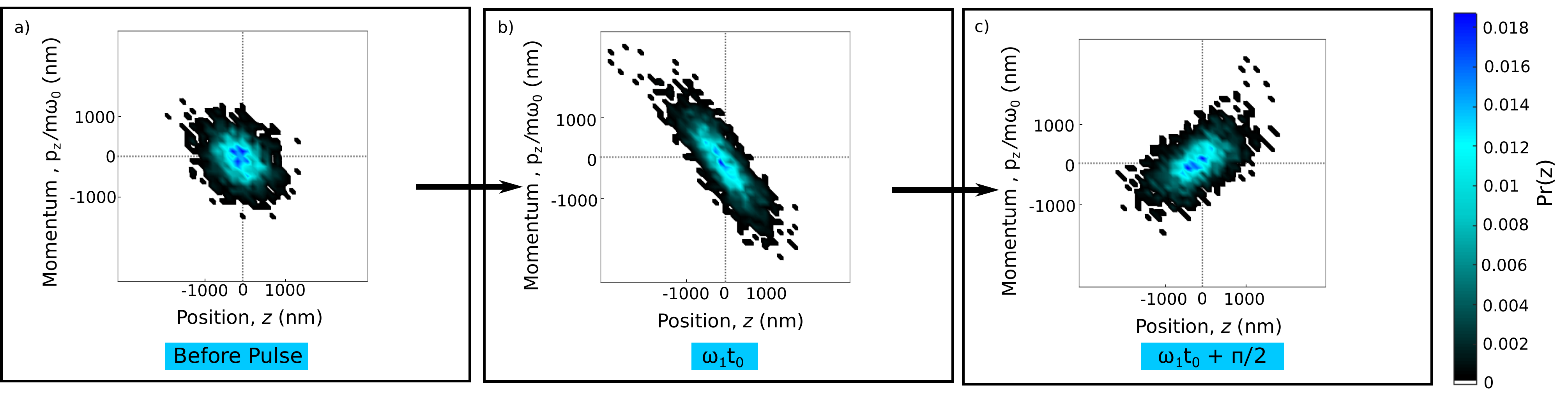}
  \caption{{Experimentally measured phase-space distributions of the
      mechanical state, before and after the squeezing pulse.} The
    average displacement of the state has been subtracted (see
    Fig.\ref{fig:one}d). ({\bf a-c}) Density plots of the phase space
    distributions for $z-$motion, at three different times, for a
    pulse duration, $\tau = 3/10 T_2$. {\bf a}) State of the particle
    motion before the pulse is applied. The former is well
    approximated by a Gaussian distribution, as typical for a thermal
    state. {\bf b}) Phase space distribution shortly after the pulse
    has been applied (time $t_0$). Note how it presents clear
    signatures of squeezing. {\bf c}) Phase space distribution at time
    $t_0+\frac{1}{4}T_1$. The squeezed state rotates in phase space
    while squeezing degrades with time, mainly due to background gas
    collisions that tend to restore the initial thermal distribution.}
\label{fig:two} 
\end{figure*} 
We trap a silica nanosphere of radius
32 nm ($\pm$5 nm) and mass of $3.1\times 10^{-19}$ kg
($\pm 1.4 \times 10^{-19}$ kg) in an optical dipole trap. The size of the
particle is evaluated from fitting a Lorentzian to the power spectral
density of the signal, as described in~\cite{Vovrosh2016} and shown in
Fig.~\ref{fig:four}b, from this the mass is obtained as well. We use a
1550 nm laser, directed into a parabolic mirror which focusses the
light to a diffraction limited spot, where the particle is trapped.
Experiments are performed in a vacuum chamber at a pressure of
$1\times10^{-1}$ mbar. In this regime, the damping of the
particle motion by random collisions with background gas is
linear in the pressure $p_{gas}$ and the related damping coefficient can
be approximated by
 \begin{equation}
  \label{eq:10a}
 \Gamma\approx 15.8\frac{r^2p_{gas}}{mv_{gas}}, 
\end{equation} 
with
$m$ and $r$ are the radius and the mass of the nanosphere,
respectively, and $v_{gas}=\sqrt{3k_BT/m_{gas}}$ is the mean thermal
velocity of the background gas of mass
$m_{gas}$~\cite{jain2016direct}. We evaluate $\Gamma = 2
\pi\times$227 Hz ($\pm 2\pi\times$9 Hz) for this experiment while the
main uncertainty in mass comes from the pressure measurement
($\sim$15 $\%$).

As shown in Fig.\ref{fig:one}a) the position of the single nanosphere
is measured using an optical homodyne method. More details about the
particle trapping and detection can be found
elsewhere~\cite{Vovrosh2016}.

A short squeezing pulse of duration $\tau$ is applied by switching
between two different trapping laser powers $P_1$ and $P_2$ (see
Fig.\ref{fig:one}b) using a free space acousto-optical modulator
(AOM). The trapping frequency is given by $\omega=\sqrt{k_{0}/m} $,
where $k_{0}={8 \alpha P}/({c \pi \epsilon_0 \rm w_{f}^{4}})$ for
motion in $z$-direction, with $\alpha$ the polarizability of the
particle, $c$ the speed of light, $\epsilon_0$ the electric field
constant and $\rm w_{f}$ the waist of the laser beam at the focal
point. The laser power can be modulated by changing the
voltage applied to the AOM; we switch between trap frequencies
$\omega_1$=2$\pi\times$112 kHz and $\omega_2$=2$\pi\times$49.3 kHz.
The timescale of the switch is determined by the AOM bandwidth, which
is more than 1 MHz and therefore much larger than both trap
frequencies. Hence, we model the switch as instantaneous. Here,
evidently, $\omega_2 < \omega_1$. The condition $\omega_2 > \omega_1$
may also be used, and it would result in squeezing of a different
quadrature. Experimentally, we found it more practical to employ the
tighter trapping potential (corresponding to $\omega_1$) most of the
time, so as to minimise the probability of losing the particle.

The same signal generator which is used to generate the squeezing
pulse triggers an oscilloscope to record a time trace of duration of
one second. The same single pulse sequence is repeated 1500 times for
the same trapped particle, while allowing for one second between the
pulses to restore the initial thermal state. The recorded time traces
initially include signals from $x, y$ and $z$ motional degrees of
freedom. However, the pulse scheme is only optimised for a single
motional frequency, namely, the one in $z$-direction which is
perpendicular to the mirror surface. This is primarily because the
$z$-motion is predominant in our detection signal. We filter the
signal around the $\omega_z$ frequency peak to extract the impact of
the pulses on the $z$-motion alone. The root mean square (rms) of the
position of the particle $z_{\rm rms}$ is used to analyse the state of
the motion, see Fig.\ref{fig:one}c). The entire experiment takes over
ten hours, during which drifts in laser power (hence in trap
frequency) may occur. Thus $\omega_2$, while known in principle, is
taken as a free parameter in the fitting model.

\begin{figure*}[t]
  \centering
 \includegraphics[width=0.95\textwidth]{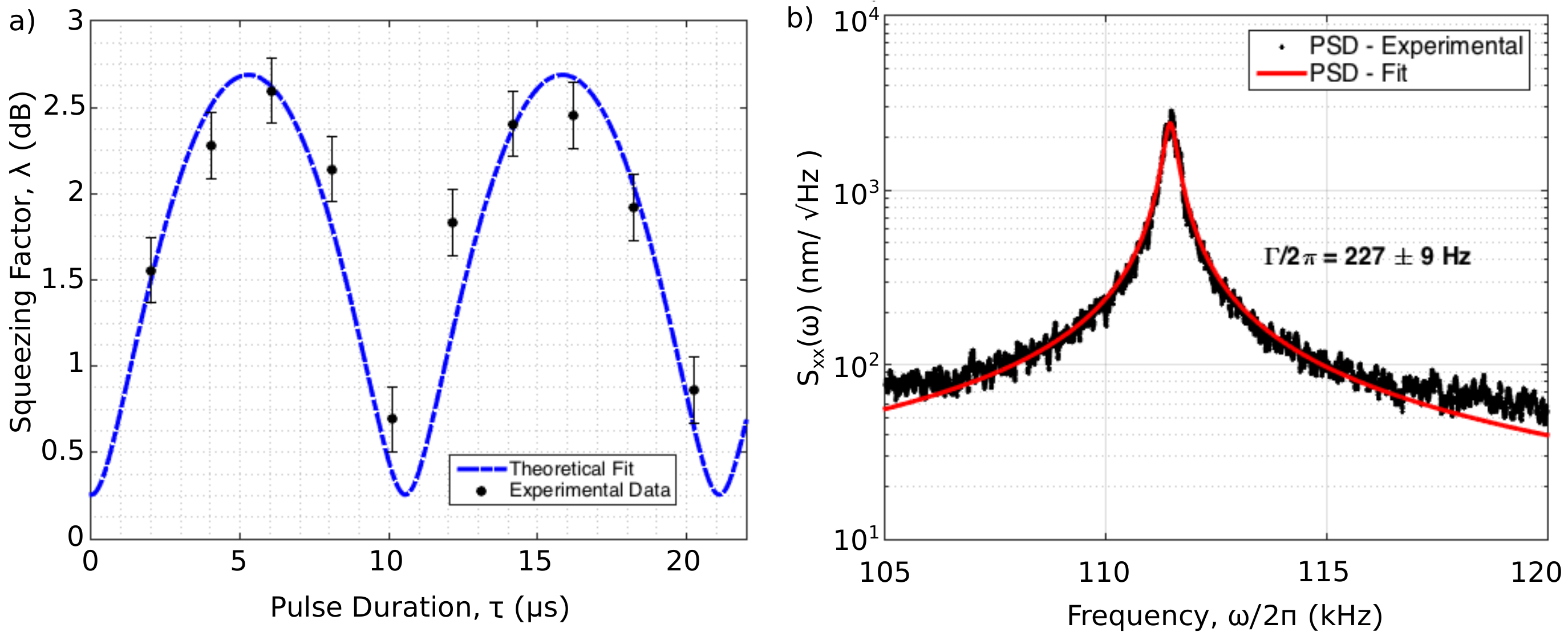}
 \caption{{\bf Quantitative analysis of the squeezing effect.} {\bf
     a}) Squeezing factor $\lambda$ as a function of pulse duration $\tau$ (measured in seconds). This is extracted by comparing the minor axes of the phase-space ellipses (see e.g. Fig~\ref{fig:two}) before and after the pulse. The theoretical fit to the data (blue line) has been done according to Eq.(10) in the supplement~\cite{supplement}. {\bf b}) Lorentzian fit to the power
   spectral density of the $z$-motion. This is used to extract the
   radius and mass of the particle, as well as, the collisional
   damping rate $\Gamma$ according to equation 9 in \cite{supplement}.}
  \label{fig:four} \end{figure*}

{\it Results--} The Fourier transform of the oscillation in the mean
position $\langle z\rangle$ - i[.e. the motion of the centre of
the thermal distribution - shows that oscillations are dominated by
frequency $\omega_1$, before and after the pulse, see
Fig.\ref{fig:one}a and~\cite{supplement}. We can thus infer that our
pulse imparts a small phase-space displacement to the particle. To
correct for this we subtract the average displacement from the data,
while we try to account for the remaining effects and experimental
imperfections through an effective dephasing model (see
supplement~\cite{supplement}). 

Initially $z_{\rm rms}$ is constant as the phase of the oscillation is
random between the 1500 individual pulse experiments. After the
squeezing pulse the motion shows damped phase-coherent
oscillations. The rms oscillation decays within about 680 $\mu$s to
690 $\mu$s, which gives a rate of
thermalization to the temperature of the background gas molecules
between 2$\pi \times $ 230 Hz and 2$\pi\times$ 234 Hz. This is in good
agreement with the value for $\Gamma$ estimated via the Lorentzian
fit.

We are operating in the classical regime, in that we observe
quadrature variances that are several orders of magnitude larger than
those in the quantum ground state. Therefore,
we may estimate the particle's momentum by simply taking the time
differential of the position measurement. In passing we note that, in
the quantum regime, our continuous measurement process would require a
more rigorous treatment~\cite{genoni2015quantum}. Applying the
described strategy to our data set we generate the phase space
distribution of the trapped particle motion. Fig.\ref{fig:two}a) shows
the distribution of the system before the pulse is applied. Such
initial distribution is nearly Gaussian, and its small asymmetry can
be attributed to the non-linear response of the position measurement
at large oscillation amplitudes~\cite{Gieseler2014}.

Immediately after the pulse we observe the typical features associated
with squeezing. It is evident that the applied pulse deforms the phase-space distribution
of the particle, which then displays the typical oblong shape of a
squeezed state: see Fig.\ref{fig:two}b). Following the pulse, the distribution rotates in phase space according to the harmonic oscillator evolution ---see for
instance Fig.\ref{fig:two}c). During such evolution the distribution
progressively relaxes back towards a thermal one; we attribute this to
thermalization via collisions with the background gas.

The measured degree of squeezing as a function of pulse duration is shown in
Fig.\ref{fig:four}a). While the theoretical prediction for the
squeezing parameter $\lambda(\tau)$ agrees qualitatively with the
experimental results, the largest squeezing factor we achieve
experimentally is 2.7 dB, lower than the expected $\lambda_{\sf
max}\simeq3.56$ dB. We can obtain a reasonable fit of the data by
assuming that the squeezing pulse is affected by some phase noise
whose strength is $\tau$-independent \cite{supplement}, and which we assume is associated with the abrupt voltage changes. In a nutshell,
this amounts to rescaling $\langle{\hat b^2}\rangle \to \eta
\langle{\hat b^2}\rangle$ at the end of the squeezing operation, where
$0\le\eta\le1$ quantifies the residual ``phase
coherence"~\cite{supplement}. For the best fit, as shown in
Fig.\ref{fig:four}a) we obtain $\omega_2=2\pi\times 47.9$ kHz ($\pm
1.55$ kHz), and $\eta= 0.73$ ($\pm 0.10$). We have assumed that other
sources of noise (e.g. thermalization with the background gas) can be neglected during the
pulse, i.e., that the motion of the particle at short timescales during the squeezing operation is affected predominantly by such phase noise.

{\it Conclusion--} The demonstrated squeezing technique could be used
for enhanced sensing and metrology based on levitated optomechanics
such as for force sensing applications~\cite{Ranjit2016} and
non-equilibrium dynamics studies~\cite{Gieseler2014}. Truly quantum squeezing may be
approached by pre-cooling the motional state ~\cite{Millen2015,
Gieseler2012, Kiesel}. Centre of mass motion temperatures of trapped
nanoparticles of below 1mK have been experimentally
demonstrated~\cite{jain2016direct, Fonseca2015, Vovrosh2016} via
parametric feedback, while alternative methods include quantum
measurement techniques~\cite{wiseman1993quantum, genoni2015quantum}
which have been successfully applied to membrane and cantilever
optomechanical devices~\cite{vanner2013cooling,
ringbauer2016generation}. Future work will include the investigation
of multiple pulses to increase the achievable levels of noise
reduction~\cite{Janszky1992}, and of methods to probe the
non-classicality of mechanical oscillators.

Finally, we would like to comment our measurement scheme, which relies
on a continuous monitoring of the particle's position. At first, this
might appear to be undesirable in the future perspective of
approaching the quantum regime, due to the well-known disturbance
induced by the quantum measurement process. Yet it was recently shown
that, if correctly accounted for, continuous monitoring may in fact
improve the achievable mechanical squeezing~\cite{genoni2015quantum}.

{\it Acknowledgements--} MR and HU would like to thank Mauro Paternostro for
the useful discussions. We acknowledge funding from the EPSRC
[EP/J014664/1], The Leverhulme Trust [RPG-2016-046], the John F
Templeton Foundation [39530] and the Foundational Questions Institute
[FQXi]. MSK thanks Danny Segal for discussions and a financial support
by the EPSRC (EP/K034480/1) and the Royal Society.

\bibliographystyle{apsrev4-1} 
\bibliography{squashing}

\newpage
\onecolumngrid
\section{Supplementary: Experimental Realisation of a Thermal Squeezed
State of Levitated Optomechanics}

\subsection{Noise model} Here we refine our theoretical
description to include phase noise during the squeezing pulse. We
obtain a reasonable fit to the data by assuming that our squeezing
operation induces some loss of coherence for mode $\hat b$ which is
independent of the pulse duration $\tau$. The idea behind this
simplification is that the most significant phase noise is generated
during the abrupt voltage changes. At the same time, all adopted pulse
lengths are short enough to prevent other ($\tau-$dependent) forms of
noise to play a significant role. It is convenient to formulate the problem in terms
of covariance matrices~\cite{serafini2005quantifying}. 

Since our state is initially thermal, and we always subtract average
displacements from the data, we may assume without loss of generality
that $\langle\hat a\rangle=\langle\hat b\rangle=0$ throughout the
dynamics. We can thus define the covariance matrix of mode $\hat a$
simply as
\begin{align}
  \sigma_a&= 
            \begin{pmatrix} 2\langle \hat X^2 \rangle&\langle \hat
              X\hat P+\hat P\hat X \rangle\\
              \langle \hat X\hat P+\hat P\hat X \rangle&2\langle P^2\rangle 
            \end{pmatrix}, 
\end{align}
where $\langle\hat A\rangle\equiv{\sf Tr}[\rho\hat A]$ for any
operator $\hat A$, and $\rho$ is the density matrix for to the
$z-$motion of our particle. Note that in our convention the covariance
matrix of the vacuum state has unit determinant. We assume an initial
thermal state of mode $a$, characterised by the covariance matrix
\begin{align}
  \sigma_a(0)&=\begin{pmatrix} 2N_1+1&0\\
    0&2N_1+1 \end{pmatrix},\\
  N_1&\equiv\frac{1}{e^{\frac{\hbar\omega_1}{k_BT}}-1}, 
\end{align}
where $T$ is the absolute temperature and $k_B$ Boltzmann's constant.
The subsequent dynamics of the covariance matrix, as induced by the
Hamiltonian $\hat H_2$, is easily determined by noticing that
$\langle\hat b^\dagger\hat b\rangle$ does not depend on time, while
$\langle\hat b(\tau)^2\rangle=\langle\hat b(0)^2\rangle
e^{-i\omega_2\tau}$.
As anticipated in the main text, we then model the loss of coherence
by simply rescaling
$\langle\hat b^2\rangle\to\eta\langle\hat b^2\rangle$, with
$0\le\eta\le 1$, while leaving $\langle\hat b^{\dagger}\hat b\rangle$
unchanged. Exploiting the known relationship between $\hat a$ and
$\hat b$, we finally obtain the following matrix elements for
$\sigma_a(\tau)$
\begin{align}
  [\sigma_a(\tau)]_{11}&=(2N_1\!+\!1)\left[\frac{1+c(\tau)}{2}+\frac{\omega_1^2}{\omega_2^2}\frac{1-c(\tau)}{2}\right],\\
  [\sigma_a(\tau)]_{22}&=(2N_1\!+\!1)\left[\frac{1+c(\tau)}{2}+\frac{\omega_2^2}{\omega_1^2}\frac{1-c(\tau)}{2}\right],\\
  [\sigma_a(\tau)]_{12}&=(2N_1\!+\!1)s(\tau)\frac{\omega_1^2-\omega_2^2}{2\omega_1\omega_2},\\
  [\sigma_a(\tau)]_{21}&=[\sigma_a(\tau)]_{12}, 
\end{align} 
where
$s(\tau)\equiv \eta\sin(2\omega_2\tau), c(\tau)\equiv
\eta\cos(2\omega_2\tau)$.
To assess the amount of squeezing that has been applied to our
mechanical oscillator, we define the quantity
\begin{equation} 
  \mu_{\sf min}(\tau)\equiv\quad\text{smallest
    eigenvalue of }\quad\sigma_a(\tau)\,, 
\end{equation}
which quantifies the variance in the squeezed quadrature. This has to
be compared with $\mu_{\sf min}(0)=2N_1+1$, the initial variance in
the oscillator quadratures at thermal equilibrium. We can thus define
a squeezing parameter (in \textit{dB} units):
\begin{equation}
  \lambda=-\frac{1}{2}10\log_{10}\left(\frac{\mu_{\sf min}(t)}{2N_1+1}\right),
\end{equation} 
where the factor of $1/2$ is due to the fact that in our work
squeezing is defined in terms of standard deviations, rather than
variances.

\section{Semiclassical description of squeezing} Since our system is a
harmonic oscillator, the classical and quantum equations of motion for
position and momentum are formally identical. Thus, a (classical)
statistical mechanics analysis of our squeezing experiment is
straightforward: it only requires to interpret $\hat X$ and $\hat
P$ as classical variables rather than quantum operators, while
expectation values should be interpreted as ensemble averages in the
classical sense, that is
$\langle f(\hat X,\hat P)\rangle=\int{\mathbb P}(x,p)\,f(x,p){\rm
  d}x{\rm{d}}p,$ for any function
$f$ of the canonical variables, ${\mathbb
  P}(x,p)$ being the joint probability density for {\it classical} position and
momentum. Differences between the two treatments would emerge in
connection with the existence of a quantum ground state, the non-commutativity of $\hat X$ and $\hat P$, and in
considering the decoherence of the system state induced by
measurements. As previously discussed, these subtleties do not play a
role in our experimental regime of high thermal excitation.

\section{Mean Position Oscillation} \label{sec:freq-time-doma}

\begin{figure}[H]
  \centering
  \includegraphics[width=0.7\textwidth]{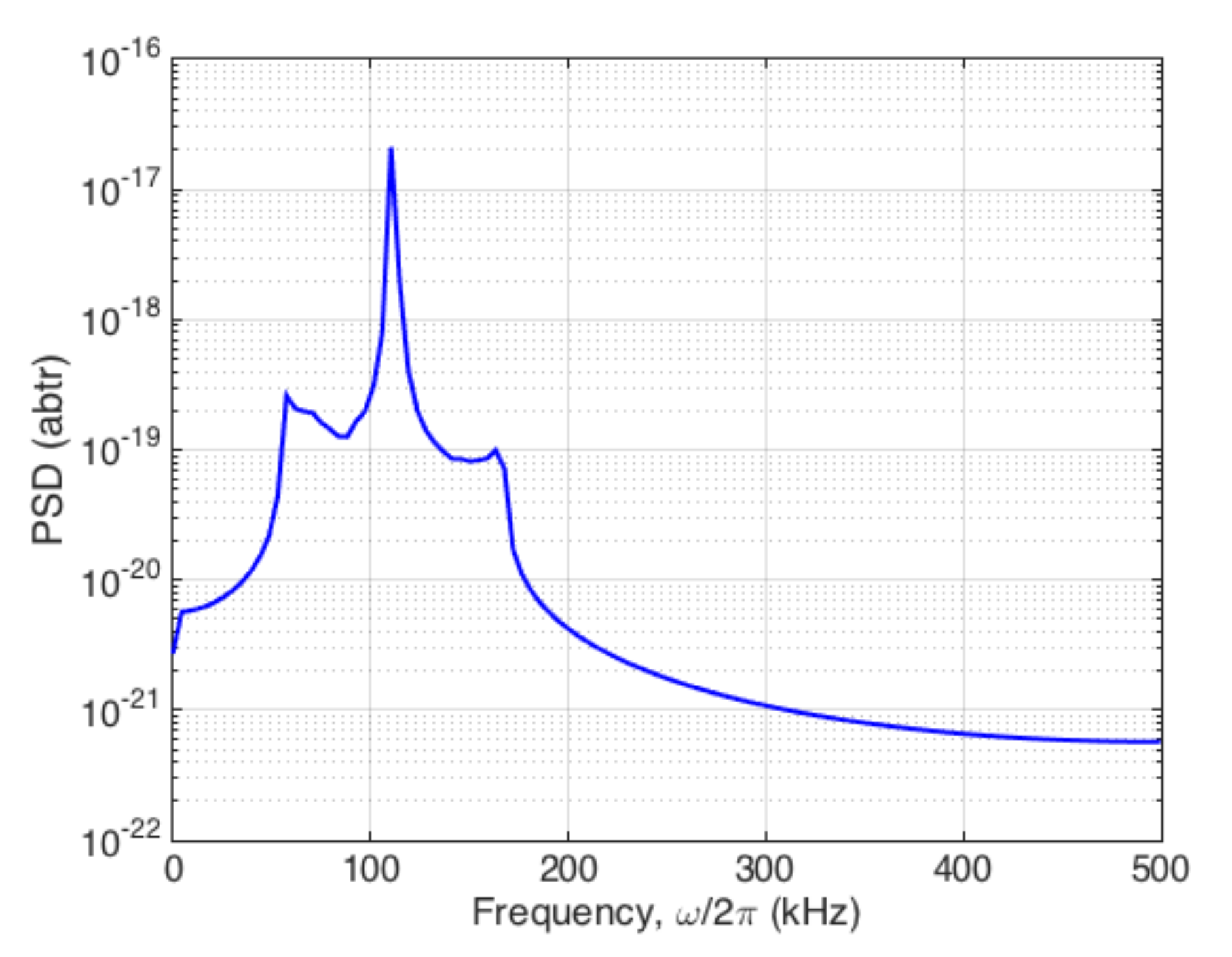}
  \caption{PSD of $\langle z \rangle$. The sharp drop off is due to
the filtering but it is clear that the most prominant frequency in the
oscillations are due to $\omega_1$.}
  \label{fig:meanzpsd} \end{figure}

As mentioned in the main text the mean position, $\langle z \rangle$
shows oscillations before and after the pulse. This refers to the
center of the ensemble oscillating in the trap. The power spectral
density is shown in figure \ref{fig:meanzpsd} of the mean position of
the particle motion. 

\end{document}